\documentclass[12pt]{article}

\usepackage{amssymb,amsmath,amsfonts,eurosym,geometry,ulem,graphicx,caption,color,setspace,sectsty,comment,footmisc,caption,pdflscape,subfigure,array,url,hyperref}

\newcolumntype{L}[1]{>{\raggedright\let\newline\\arraybackslash\hspace{0pt}}m{#1}}
\newcolumntype{C}[1]{>{\centering\let\newline\\arraybackslash\hspace{0pt}}m{#1}}
\newcolumntype{R}[1]{>{\raggedleft\let\newline\\arraybackslash\hspace{0pt}}m{#1}}

\begin{document}

\begin{titlepage}
\title{Is the NFL’s franchise tag fair to players?}
\author{Darwin Zhou\thanks{I sincerely thank Ms. Sara Shreve-Price and Ms. Julia Nickles for providing me with the knowledge and help to write this paper.}}
\date{\today}
\maketitle
\begin{abstract}
\noindent There has been a consistent criticism over the past decade of the NFL franchise tag’s monetary limitations due to its biased institutions in favor of the team rather than the player. But the question whether the NFL’s franchise tag is fair or unfair to players has never been systematically studied. In this paper, I investigate the effects of NFL players’ contract extensions when on a franchise tag compared to when they are not and analyze them through statistical and economic lens. Through my research, I find that indeed the current franchise tag designation is unfair to players when it comes to contract extension. I then propose a solution to remedy this unfairness, that is, removing the opportunity to franchise tag players for multiple years, and adding an option for the player to either test free agency but receive zero pay until they settle on a contract (the team can also match the offer) or sign the franchise tag, to provide more flexibility for the player and the team. \\
\vspace{0in}
\vspace{0in}

\bigskip
\end{abstract}
\setcounter{page}{0}
\thispagestyle{empty}
\end{titlepage}
\pagebreak \newpage

\doublespacing

\section{Introduction} \label{sec:introduction}
It is safe to say that sports are an integral part of many Americans’ lives. Even with a global pandemic, sports viewership has been steadily increasing in America. \textit{It is estimated that by 2025, an average of 90 million Americans will be watching live sports at least once per month}~\footnote{https://www.statista.com/statistics/1127341/live-sport-viewership/}. 
Out of all the different professional sports leagues, the NFL (National Football League) accounts for the majority of viewership, and has seen a nonstop growth every year in its audience. \textit{A projected massive total of 168 million people watched the NFL during the course of its season}; the 2021 NFL regular season averaged 17.1 million viewers, a 10\% increase from the 2020 NFL regular season~\footnote{https://bit.ly/3PgL11q};
and to top it all off, what’s considered the world’s most viewed sporting event, \textit{the Super Bowl, estimated a total of 112.3 million viewers}~\footnote{https://bit.ly/3PhzdvN}, 
which meant approximately one-third of the entire US population spent their Sunday afternoon in front of a TV screen to tune in to watch football. With viewership booming and revenue coming in from every direction, \textit{the NFL is named the most profitable sports league in the world}~\footnote{https://bit.ly/3Pm76LQ}. 
The profound success of the league allows its \textbf{salary cap}, which is the upper limit on the amount of money players on a team can be paid, \textit{to increase from \$182.5 million during the 2020 COVID year to \$208.2 million}~\footnote{https://bit.ly/3Ql82Bv}, 
and this number is projected to continue to grow. 

One of the most intriguing parts of the NFL season is the off-season, which is a period of exhilaration and anticipation for fans, as they will get to witness players in new places or stay with their current teams. Just this season, we have seen record-breaking contracts all across the league, the most notable being the Cleveland Browns paying quarterback Deshaun Watson \$230 million all in guaranteed money (the largest guaranteed contract in NFL history)~\footnote{https://bit.ly/3QhN3PX}.
\textit{Something important to note about NFL contracts is that the money is only partially guaranteed}, and instead contract extensions provide the max amount (including incentives) that a player could make based on how well they perform, making the fully guaranteed that Deshaun Watson earns even more incredible. This raises a question of if all NFL teams would be providing their superstar players large contract extensions.

The answer is no, and it is all because of a particular designation that the NFL founded almost 20 years ago back in 1993, called the \textbf{franchise tag}, which many NFL players now dread to be signed under. In short, the franchise tag allows a team to sign a player who they consider to be a “franchise” player - an integral part of the team - to a fixed, one year deal with an appropriate amount of guaranteed money depending on the player’s position. At the first glance, the franchise tag appears to be mutually beneficial to the team and the player. The team gets to retain a key player that largely impacts the team’s performance while not breaking the bank, and the player gets a decent amount of guaranteed money. However, in recent years, the franchise tag has seemed to become \textit{lopsided towards the team and less towards the player}, leading to an outbreak of criticism by many as people believe that the franchise tag has become an exploitative tool for teams. 

\textit{It is important to also know that the franchise tag can be used multiple years up to three times}, but it is nowhere as common as the one-year franchise tag. Although the amount of guaranteed money does go up, it is not by a substantial amount. The backlash for teams tagging a player multiple times has been well circulated, and has been most well stated by Seattle Seahawks All-Pro offensive tackle Walter Jones, one of two players ever to be tagged for three consecutive years~\footnote{https://www.fieldgulls.com/2007/4/6/204056/2795}:
“Maybe when it [franchise tag] was invented, it was good…teams tell you how much you should be flattered that they think enough of you to make you their franchise guy…but it's not a thriller…it's a killer watching all the deals get signed with huge bonuses and you're not getting the big money upfront. It’s a lousy system." 

These criticisms raise an intriguing question: is the NFL's franchise tag fair to players? To the best of my knowledge, this question has been systematically studied. In this paper, I make an attempt to answer this question. My main contributions are to research the logistics behind this criticism, that is, players on a franchise tag obtaining smaller contract extensions than players not on a franchise tag, through statistical and economic lens and to offer a way of bridging the socioeconomic gap~\cite{blair2011sports} between the team and players on the franchise tag. 

The paper proceeds as follows. Section \ref{sec:data} discusses the data and methods used. Section \ref{sec:result} presents the results. Section \ref{sec:discussion} presents related discussions to conclude the paper.


\section{Data and Methods} \label{sec:data}
\subsection{Data}

\begin{figure}[h]
\includegraphics[width = \textwidth]{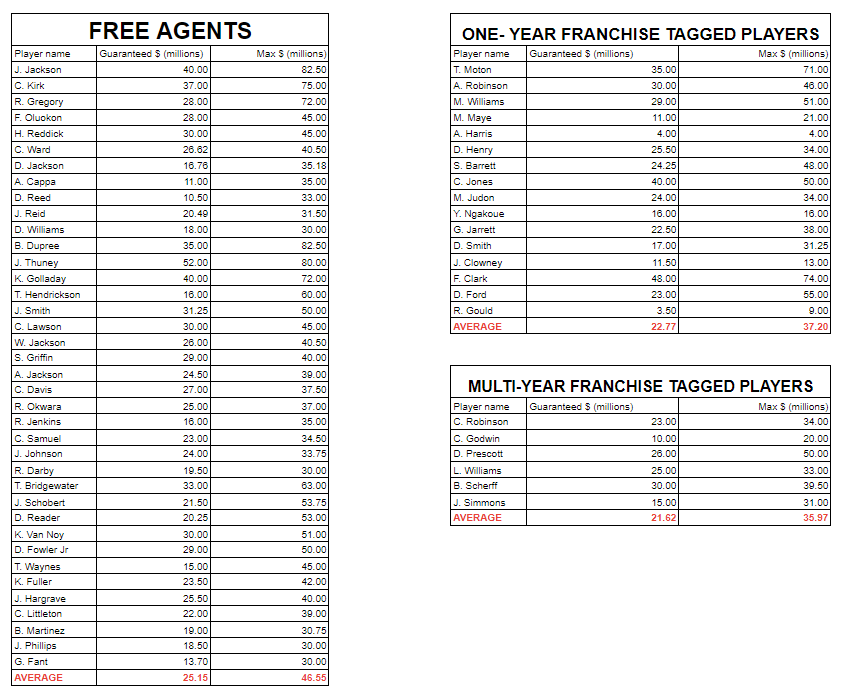}
\centering
\caption{The tabulated contract extension information of free agents, one-year franchise tag players and multi-year franchise tag players for the 2020-2022 NFL off-seasons.}
\label{fig:table}
\end{figure}

The common criticism of the franchise tag has been the fact that players under the designation have little to no negotiation powers for their contract and will now have a smaller chance of signing a large contract extension the following season, regardless they are a free agent or not. With the knowledge that an NFL player’s prime is considerably shorter than other sports due to the violent nature of the game, \textit{ranging from 25-29 years old}, and the fact that the average contract extension (the contract after a player’s rookie contract) takes place around that age range as well, it is imperative for players to secure what could be the money they live the rest of their life on once their rookie contract is over. Teams are inclined to have higher expectations for franchise tagged players, so they will be met with an excessive amount of pressure. In addition to the volatile nature of football, whether injuries or performance-wise, if anything negative happens to the franchise tagged player, it is almost certain that their free agency stock will drop to extremes where they may even be out of the league the following year. Even though players can \textbf{“hold out”} (a term used by football fans for players who voluntarily decide to sit out games and practices for the purpose of seeking a more expensive contract) when under the franchise tag, they paint themselves as a self-centered agitator, which discourages teams from offering them a contract. Figure \ref{fig:table} shows a chart of the amount of money players earn under the franchise tag. \textit{Although it is fully guaranteed, it is nowhere near the hundreds of millions of potential money players could be earning through free agency.} The 2022 NFL off-season was record-breaking, with 40 players earning 50+ million dollar contract extensions, the largest amount in an NFL off-season ever, equating to a total of about 2.3 billion dollars in guaranteed money.

To support my hypothesis of players on a franchise tag obtaining smaller contract extensions than players not on a franchise tag, I separate my contract extension data into three study groups: (1) players who were free agents,(2) players under a one-year franchise tag, and (3) players under a multi-year franchise tag. The criteria I use for obtaining my data were as follows:

\begin{itemize}	
\item They all had to be players from the same off-season.

\item They all have to have signed their contracts within the 25-29 age range - the prime age for an NFL player - when most NFL players sign their first contract extension.

\item The players in the free agent group all had to be “superstar” or above average players, meaning they either have made a Pro Bowl or All-Pro selection or have garnered consistent production, since franchise-tagged players tend to be among the top 10\% at their position. 
\end{itemize}

I collect my data through reliable and in-depth NFL statistical websites, such as Spotrac~\footnote{https://www.spotrac.com/}, a website dedicated to contracts and the breakdowns of each one, and Pro Football Reference~\footnote{https://www.pro-football-reference.com/}, a forum that provides player statistics and information. \textit{My data consist of 38 free agents, 16 one-year franchise tagged players, and 6 multi-year franchise tagged players,} sorted by three categories of player name, guaranteed money made from signing the contract, and maximum amount of money made from signing the contract. Figure \ref{fig:table} shows the actual data I curate for analysis for the three most recent off-seasons from 2020 to 2022, in order to provide a big enough sample size to conduct a significance test and make inferences on. 

\subsection{Methods}

I conduct my analysis through the following four means.

1) \textbf{Table comparison}. I put the curated data into tables that compare guaranteed money and max amount of money earned from the contract for the three study groups, where I calculate the average amount for each circumstance as well as a baseline comparison. 

2) \textbf{Box plot chart}~\cite{book}. 
I create two box and whisker plot charts to help better visually represent my data. For the plots, I utilize the minimum, first quartile (25th percentile), third quartile (75th percentile), and maximum values from the respective datasets. Tables 3 and 4 show my box and whisker plots. 

3) \textbf{Statistical significance test}~\cite{book}.
I conduct six two sample t-tests for the difference between two means comparing free agents and franchise tagged (one year and multi-year combined) players in terms of their guaranteed money and maximum amount of money they earn on their contract extensions through two pooled data tests and four non-pooled data tests. 

Mathematically, the six tests of the same nature are defined as follows.
\begin{itemize}
    \item{Test 1} – Guaranteed money of free agent ($\mu_1$) vs (pooled) franchise tagged player ($\mu_2$) 
    \begin{equation} \label{eqn:test1}
H_0: \mu_1 = \mu_2 ~~ vs ~~  H_a: \mu_1 > \mu_2
\end{equation}
where\\
$\mu_1$: population mean of free agents contract extension guaranteed money,\\
$\mu_2$: population mean of franchise tagged players (one + multi year) contract extension guaranteed money.

\item{Test 2}  – Guaranteed money of free agent ($\mu_1$) vs one-year franchise tagged player ($\mu_2$).

\item{Test 3}  – Guaranteed money of free agent ($\mu_1$) vs multi-year franchise tagged player ($\mu_2$).

\item{Test 4} – Maximum money of free agent ($\mu_1$) vs (pooled) franchise tagged player ($\mu_2$).

\item{Test 5} – Maximum money of free agent ($\mu_1$) vs one-year franchise tagged player ($\mu_2$);

\item{Test 6} – Guaranteed money of free agent ($\mu_1$) vs multi-year franchise tagged play ($\mu_2$).

\end{itemize}

4) \textbf{Payoff matrix}~\cite{econ}. 
I create two payoff matrices that showcase the projected outcomes of the actions that the player and team takes before my proposed solution and after my proposed solution. 

\section{Results} \label{sec:result}
As shown in Figure \ref{fig:table}, \textit{there is a clear pattern that the more years on the franchise tag, the guaranteed and total potential money a player will make gets lower.} For all the free agent players, the total average guaranteed and total average max money was 25.15 million and 46.55 million dollars, respectively. For the franchise tagged players, the total average guaranteed and total average max money from contract extension was 22.77 million and 37.2 million dollars, respectively. For the multi-year franchise tagged players, the total average guaranteed and total average max money from contract extension was 21.62 million dollars and 35.97 million dollars, respectively. 

\begin{figure}[h]
\includegraphics[width =0.45 \textwidth]{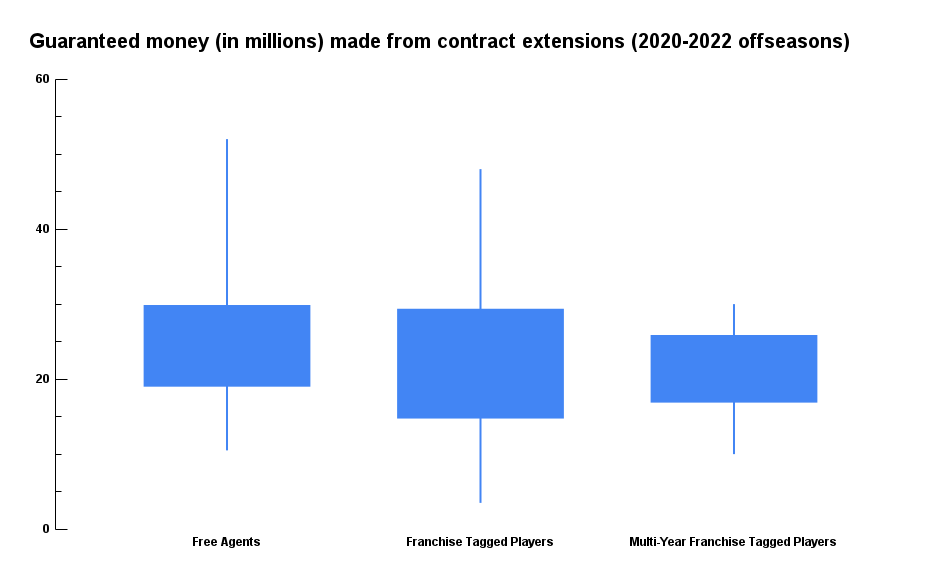}(a)
\includegraphics[width =0.45 \textwidth]{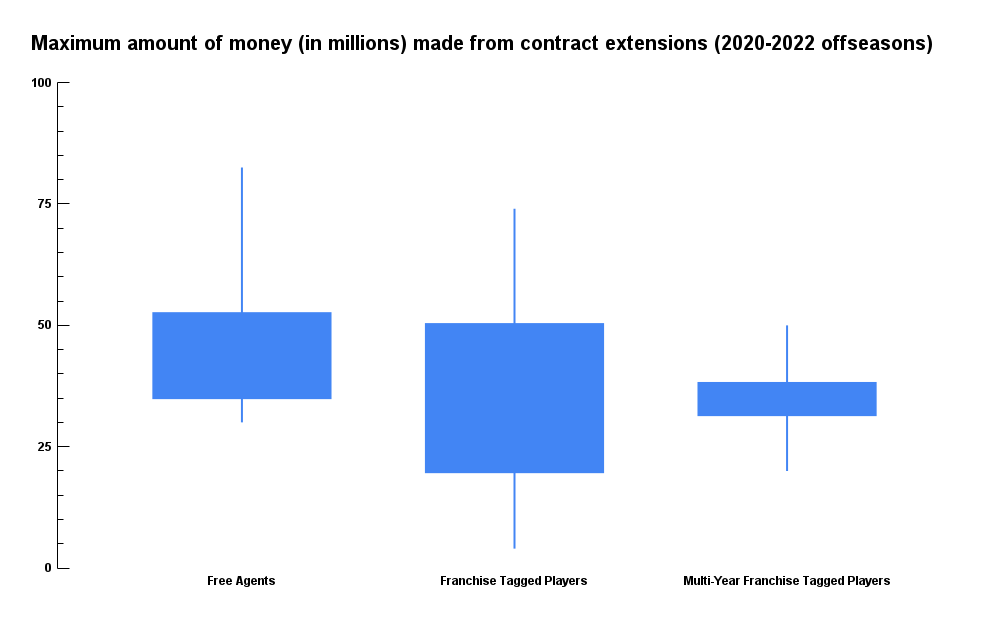}(b)
\centering
\caption{The box and whisker plots for (a) the guaranteed money and (b) the maximum money made from contract extensions for free agents, franchise tagged players, and multi-year franchise tagged players.}
\label{fig:box}
\end{figure}

When comparing the box and whisker plots as in Figure \ref{fig:box}, for both guaranteed money and max amount of money earned from contract extensions, \textit{the variance of the multi-year franchise-tagged players is significantly less than the variance of the free agents and franchise-tagged players.} When examining the median of free agents (24.75, 40.50), it is also higher than both franchise tagged players (23.5, 36), and multi-year franchise tagged players (24, 33.5) for both guaranteed and max amount of money earned from contract extensions. Overall, these statistics prove that there is a huge advantage for free agents signing contract extensions when compared to players on any type of franchise tag. 

Through my six significance tests, I obtain p-values of 0.153, 0.252, 0.161, 0.012, 0.067, and 0.016. \textit{These p-values mostly indicate that the relationship between free agents and franchise-tagged players (one and multi-year) is statistically significant}, meaning that its likely that the massive difference in money made from contract extensions is actually real rather than by chance. 

However, numbers merely do not show enough. Let us examine the selected players’ situations closer for the most recent off-season, 2022. When looking at the franchise tagged players, three out of the four of them had problems. Allen Robinson had a career-worst season with the Chicago Bears, Chris Godwin tore his ACL, and Marcus Maye tore his Achilles. As for the multi-year franchise tagged players, two out of the four of them had issues. Dak Prescott suffered a near career-ending ankle injury, and Brandon Schreff tore his MCL. It is safe to presume that after seeing their superstar player go down with injury, the respective teams do not feel as high of an incentive to pay them top dollar money. Combining that with the fact that the more years under the franchise tag, the older the player gets within their prime, meaning that the player's performance inevitably declines, teams is more hesitant to sign an older player coming off an injury than just browsing free agency for younger and more productive assets. 

\begin{figure}[h]
\includegraphics[width = \textwidth]{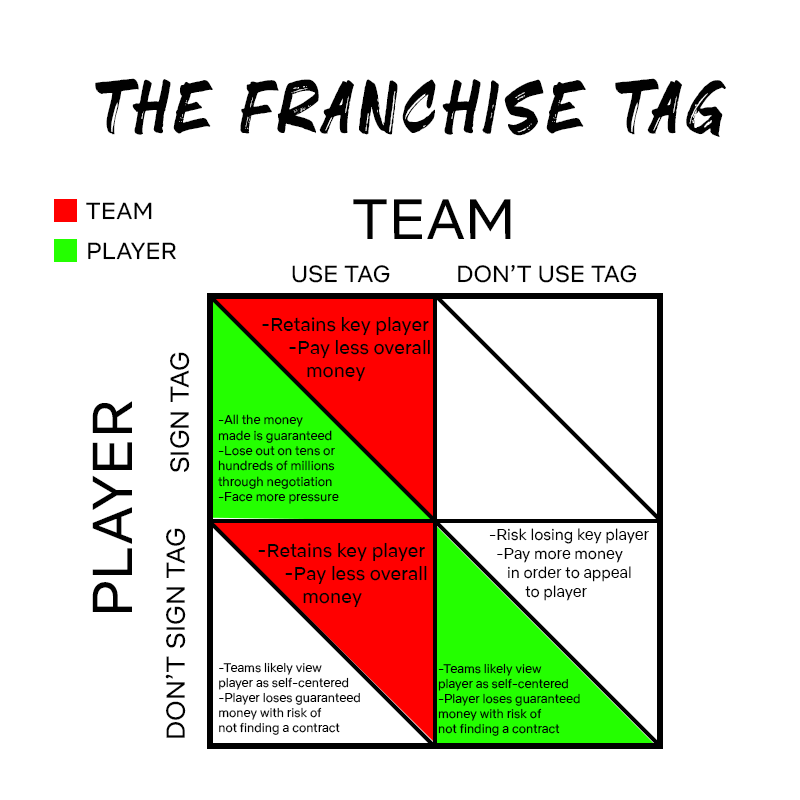}
\centering
\caption{The game-theoretic aspects for the tag player and team pre-solution.}
\label{fig:pre-solution}
\end{figure}

Figure \ref{fig:pre-solution} shows game-theoretic aspects by listing the simultaneous choices of the player and the team pre-solution. \textit{There is a clear dominant strategy for the team to use the franchise tag, as the benefits massively outweigh the risks, but the player loses in every scenario.} The player is either viewed as someone that is not team-friendly and loses out on guaranteed money through holding out by not playing on the franchise tag, or faces the massive risk of earning less potential money whilst dealing with immense pressure by playing on the franchise tag. 

\section{Discussions} \label{sec:discussion}

As one-sided as the franchise tag is, it should not be removed from the NFL. \textit{Is there a way that we could reach a win-win, mutually beneficial situation between the team and the player which is always the intended purpose of the franchise tag?} In fact, there are two things the NFL should implement to make the franchise tag more fair. 

First, the NFL needs to completely remove the multi-year franchise tag and make the franchise tag simply one year long. When looking at the past three seasons, the amount of multi-year franchise tagged players is so minimal (less than half of franchise tagged players), and further they make less average guaranteed and max money from contract extensions compared to the other groups. The low p-values of 0.161 and 0.016 provide evidence that the massive difference in money made contract extensions is in fact real. The range of money players make from the multi-year franchise tag as shown by the box and whisker plot is actually substantially lower than the other two study groups. The franchise tag itself is already extremely unfair to the player since it drastically reduces the amount of money they could be making while supplying them with added pressure, so if a player has to go through the same process for two or three years, it means they lose out on tens, even hundreds of millions once they reach their contract extensions. 

Second, the NFL should add a team and player option to the franchise tag to bring the mutuality aspect that the designation lacks. For the player, other than signing the tag, they are given a second option where they are allowed to test free agency and seek a contract from other teams, however, they will not be eligible to play in games and therefore will be without pay for the entire process until they settle on an agreement. For the team, if the player does end up finding a contract, they have the opportunity to match the offer and keep the player. By using these choices, it’ll actually give the player a fair chance of seeking a large contract extension which the current franchise tag does not offer at all, but it will still be fair for the team since they won’t have to pay the player during the process as they won’t be playing in games. Once the player does go through the lengthy process of negotiating a contract, then the team still has a chance to keep their star player, but they’ll have to match the price. This maintains the strategic element of using such a valuable designation like the franchise tag, but in a fair and non exploitative way. The team will have to truly decipher whether or not the player is worth keeping, instead of just keeping them for the sake of saving millions in cap space. 

\begin{figure}[h]
\includegraphics[width = \textwidth]{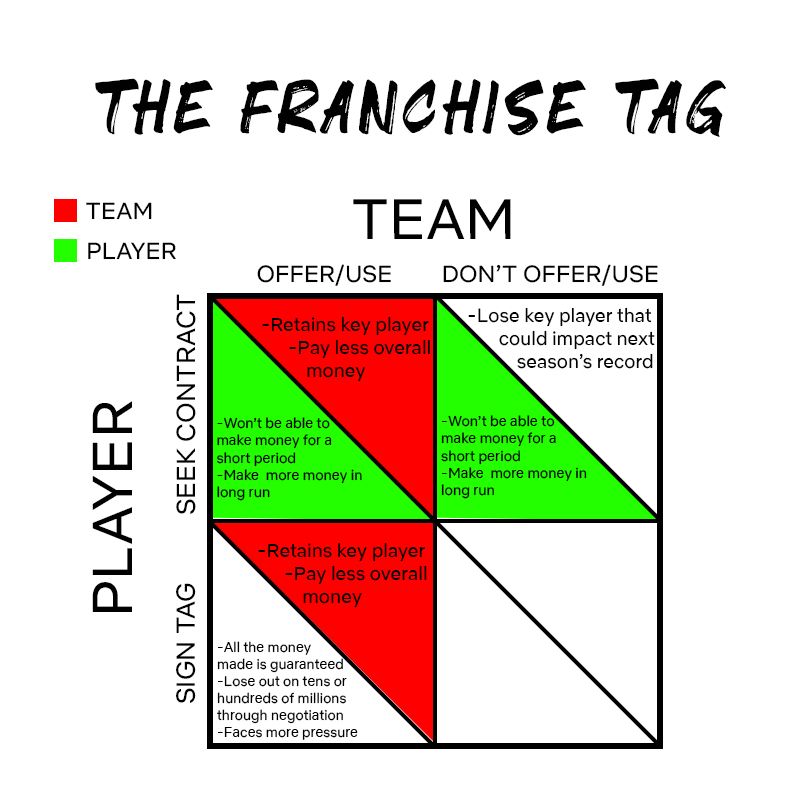}
\centering
\caption{The game-theoretic aspects for the tag player and team post-solution.} \label{fig:solution}
\end{figure}

Figure \ref{fig:solution} shows game-theoretic aspects by listing the simultaneous choices of the player and the team post-solution. \textit{Both the team and player have a dominant strategy} - for the team it is using the tag/offering a contract - for the player it is seeking a contract extension. The Nash equilibrium would be the team using the tag/offering a contract and the player seeking a contract extension. The solution fixes the problem of the franchise tag not being mutually beneficial. 

Underpaying labor workers and wealth inequality is a huge problem in today’s society, so we definitely do not want to witness players, who risk their entire bodies as a career, get burdened as well and lose out on hundreds of millions of dollars. Hopefully, with the above two suggestions, the NFL can utilize the franchise tag the way it’s supposed to: as a synergistic nomination rather than a one-sided tool.

\bibliographystyle{plain}
\bibliography{references}

\newpage
\section*{Figures} \label{sec:figures}

\begin{figure}[h]
\includegraphics[width = \textwidth]{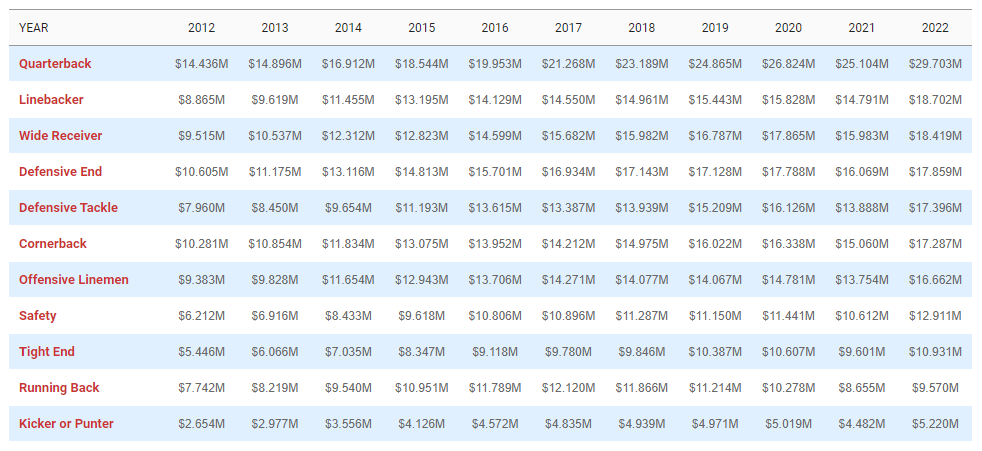}
\centering
\caption{The amount of money players earn under the franchise tag. Source: Spotrac}
\end{figure}

\end{document}